\documentclass[epj,referee]{svjour}
\usepackage{graphics}
\usepackage[dvips]{color}
\usepackage{amsmath,amssymb}

\begin{document}
\title{Robustness of correlated networks against propagating attacks}
\author{Takehisa Hasegawa\inst{1} \and Keita Konno\inst{2} \and Koji Nemoto\inst{2}
}                     
%
%
\institute{Graduate School of Information Sciences, Tohoku University, 6-3-09, Aramaki-Aza-Aoba, Sendai, 980-8579, JAPAN. 
\and Department of Physics, Graduate School of Science,Hokkaido University, Kita 10-jo Nisi 8-tyome, Sapporo, JAPAN.}
\date{Received: date / Revised version: date}
%
\abstract{
We investigate robustness of correlated networks against propagating attacks 
modeled by a susceptible-infected-removed model.
By Monte-Carlo simulations, we numerically determine 
the first critical infection rate, 
above which a global outbreak of disease occurs, 
and the second critical infection rate, 
above which disease disintegrates the network.
Our result shows that correlated networks are robust compared to the uncorrelated ones, 
regardless of whether they are assortative or disassortative, when a fraction of infected nodes in an initial state is not too large. 
For large initial fraction, disassortative network becomes fragile while assortative network holds robustness.
This behavior is related to the layered network structure inevitably generated by a rewiring procedure we adopt to realize correlated networks.
} 
\maketitle
\section{Introduction}

One of the most important properties of complex networks, such as the WWW, the Internet, social and biological networks, is 
robustness against failures or intentional attacks
~\cite{albert2002statistical,newman2003structure,boccaletti2006report,dorogovtsev2008critical,barrat2008dynamical,newman2010networks}.
Albert {\it et al}.\ numerically studied the robustness of networks against 
random failure, where nodes are sequentially removed with equal probability,
and intentional attack, 
where hubs (nodes with large degrees) are preferentially removed~\cite{albert_error_2000}.
When a network has the scale-free (SF) degree distribution $p_k \propto k^{-\gamma}$,
where $k$ is degree and $p_k$ is the fraction of nodes with degree $k$, 
the network having $\gamma \le 3$
is highly robust against the random failure: 
you have to remove almost all nodes to disintegrate the network.
However, SF networks are fragile to the intentional attack:
the network is destroyed if a small fraction of hubs are removed.
Later, this observation has been analytically supported~\cite{callaway2000network,cohen2000resilience,cohen2001PRL}.
The robustness of networks against other percolation-like processes such as
the betweenness-based attack~\cite{holme2002attack} and
degree-weighted attacks~\cite{gallos_stability_2005} have also been studied.

Other attacks to networks may occur as propagating processes such as computer viruses do
~\cite{kephart1991directed,kephart1993measuring,Pastor-Satorras2001epidemicPRL}.
Let us consider the susceptible-infected-removed (SIR) model with infection rate $\lambda$ (and recovery rate $\mu=1$) on a network. 
The system has, in general, two critical infection rates $\lambda_{c1}$ and $\lambda_{c2}$ \cite{Newman-Threshold-2005PRL,hasegawa2011robustness}.
Above the first critical infection rate $\lambda_{c1}$, a global outbreak of disease occurs. 
In a global outbreak, a fraction of the nodes become infected and eventually removed.
The remaining network of susceptible nodes, however, may survive as a giant component
up to the second critical infection rate $\lambda_{c2}$, above which the network is finally disintegrated.
Thus we here adopt $\lambda_{c2}$ as a measure of robustness against propagating attacks
\cite{hasegawa2011robustness}. 
Note that $\lambda_{c2}$ is in principle equal to or larger than $\lambda_{c1}$.

In~\cite{moreno2002epidemic}, the first critical infection rate $\lambda_{c1}$ is approximately derived.
In particular, any positive infection rate induces a global outbreak in SF networks having $\gamma \le 3$, meaning $\lambda_{c1}=0$.
Newman studied the SIR model on uncorrelated networks in terms of transmissibility
to show that $\lambda_{c2}>0$ even for $\gamma \le 3$~\cite{Newman-Threshold-2005PRL}. 
Hasegawa and Masuda extended his analysis to evaluate the robustness $\lambda_{c2}$ in some vaccinated networks~\cite{hasegawa2011robustness}.

Real networks often have some degree correlations
~\cite{newman2002assortative,newman2003mixing}.
A network is said to be assortative when nodes with similar degrees tend to connect with 
each other.
A network is called disassortative when nodes with high degrees tend to connect to nodes with low degrees.
Empirical data indicates that social networks are likely to be assortative, while technological networks and biological networks to be disassortative
~\cite{newman2002assortative,newman2003mixing,newman2003structure}.
Degree correlations affect dynamics on networks.
Previous studies have reported the effects of the degree correlations on various dynamics, e.g., 
percolation \cite{valdez2011effect,goltsev2008percolation}, target attack \cite{shiraki2010cavity,tanizawa2011robustness}, 
Ising model \cite{menche2011sequences}, susceptible-infected-susceptible model \cite{boguna2003absence}, information spreading \cite{schlapfer2012decelerated}, etc.
As for the SIR model, Vazquez and Moreno showed that the increase of the assortativity makes easier 
to induce a global outbreak \cite{vazquez2003resilience}.
However, we do not know the effect of the degree correlation of networks on 
the robustness against such propagating attack modeled by the SIR model.

In this paper, we perform a Monte-Carlo simulation for the SIR model on networks having a degree correlation.
We numerically determine $\lambda_{c1}$ and $\lambda_{c2}$ to show that correlated network is robust compared to the uncorrelated one 
regardless of whether it is assortative or disassortative, when a fraction of infected nodes in an initial state is not too large. 
For large initial fraction, disassortative network becomes fragile while assortative network holds robustness.
This behavior is related to the layered network structure 
inevitably generated by a rewiring procedure we adopt to realize correlated networks~\cite{menche2011sequences,xulvi2004reshuffling}.


\section{Model}

We consider the SIR model on a given network with $N$ nodes.
Each node takes one of the three states: susceptible, infected, or removed.
A fraction $p_{\rm seed}$ of the nodes are initially infected and other nodes are susceptible.
Susceptible nodes get infected at a rate proportional to the number of infected neighbors:
the susceptible node gets infected with probability $\lambda \Delta t$ within a 
short time $\Delta t$ when it is adjacent to an infected node. 
An infected node becomes removed, irrespective of the neighbors' states, at a unit rate $\mu=1$,
i.e., with probability $\Delta t$ within short time $\Delta t$.

In the final state, each node takes either susceptible or removed state.
We call the connected components of removed nodes and susceptible nodes the infected networks and the residual networks, respectively.
Then, the critical infection rates $\lambda_{c1}$ and $\lambda_{c2}$ are given 
in terms of percolation transition.
We define $R_{\rm max}(N)$ and $S_{\rm max}(N)$ to be the mean largest component sizes 
of the infected networks and residual networks, respectively.
Noting that an infection quickly dies out (spreads globally) for $\lambda<\lambda_{c1}(>\lambda_{c1})$ and 
the residual networks include a giant component (consist of only finite components) for $\lambda<\lambda_{c2}(>\lambda_{c2})$,  we expect that
$R_{\rm max}(N)$ and $S_{\rm max}(N)$ behave as
\begin{eqnarray}
\{R_{\rm max}, S_{\rm max} \} \propto
{\Biggl\{}
\begin{array}{ccl}
\{ {\rm const.}, N \} & (\lambda < \lambda_{c1}) & \\
\{ N, N \} & (\lambda_{c1} <\lambda< \lambda_{c2}) & \\
\{ N, {\rm const.} \} & (\lambda> \lambda_{c2}) &
\end{array}. \label{Phase1}
\end{eqnarray}

In the following sections, we perform Monte-Carlo simulations 
for the SIR model on the uncorrelated and correlated networks. 
Here we adopt the following degree distributions: 
an exponential degree distribution, 
\begin{eqnarray}
p_k &=& \frac{1}{3}\left( \frac{2}{3} \right)^{k-2}, \label{degExp}
\end{eqnarray}
and a SF degree distribution with $\gamma=3$, 
\begin{eqnarray}
p_k &= & \frac{12}{k(k+1)(k+2)} \propto k^{-3}. \label{degPow}
\end{eqnarray}
Note that those networks have the same average degree $\langle k \rangle=4$.

An uncorrelated network having the above distributions can be generated by the configuration model \cite{newman2003structure}.
The network so obtained is then randomly rewired so as to have a correlation without changing
the degree distribution.
We adopt the following scheme proposed in \cite{menche2011sequences,xulvi2004reshuffling}: 
(a) Select randomly two edges and look up the degrees of four nodes linked by these edges.
(b) Rewire the two edges, with probability $p_{\rm rewire}$, in a way that 
one edge links two nodes having the higher degrees and the other links the remaining two nodes,  and otherwise rewire them randomly.
(c) Repeat (a) and (b) until the degree correlation (see below) reaches a stationary value.
To generate a disassortative network, we replace the step (b) with the following step:
(b)$^\prime$ Rewire the two edges, with probability $p_{\rm rewire}$, in a way that
one edge links two nodes having the highest and lowest degrees and
the other links the remaining two nodes, and otherwise rewire them randomly.

Here we adopt the following quantity for a measure of the degree correlation~\cite{xulvi2004reshuffling}:
\begin{eqnarray}
 \mathcal{A}=\frac{\sum_k\mathcal{E}_{kk} -
\sum_k\mathcal{E}^{r}_{kk}}{1-\sum_{k}\mathcal{E}^r_{kk}},
\end{eqnarray}
where $\mathcal{E}_{kk}$ is the probability that both ends of a randomly chosen edge have degree $k$, 
and $\mathcal{E}^{r}_{kk}$ is that of the uncorrelated networks (with the same degree distribution).
Uncorrelated networks take $\mathcal{A}=0$, 
and a network is regarded as assortative (disassortative) when $\mathcal{A} > 0 ( < 0)$. 
After the above process, the mean value of $\mathcal{A}$ takes a $p_{\rm rewire}$-dependent value.
Choosing $p_{\rm rewire}$ appropriately, we have a network with a desirable degree correlation.


\begin{figure}
\centering
\resizebox{0.7\columnwidth}{!}{%
  \includegraphics{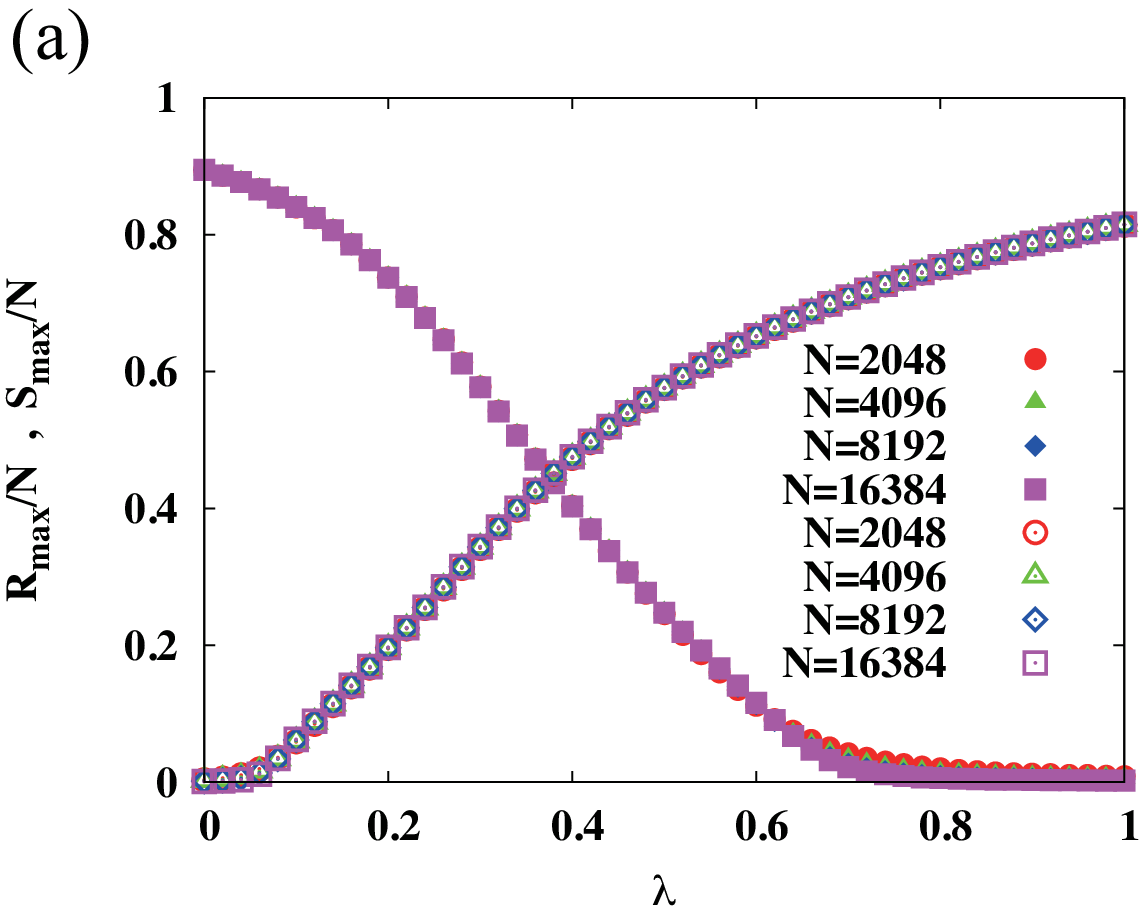}
  }
\resizebox{0.7\columnwidth}{!}{%
  \includegraphics{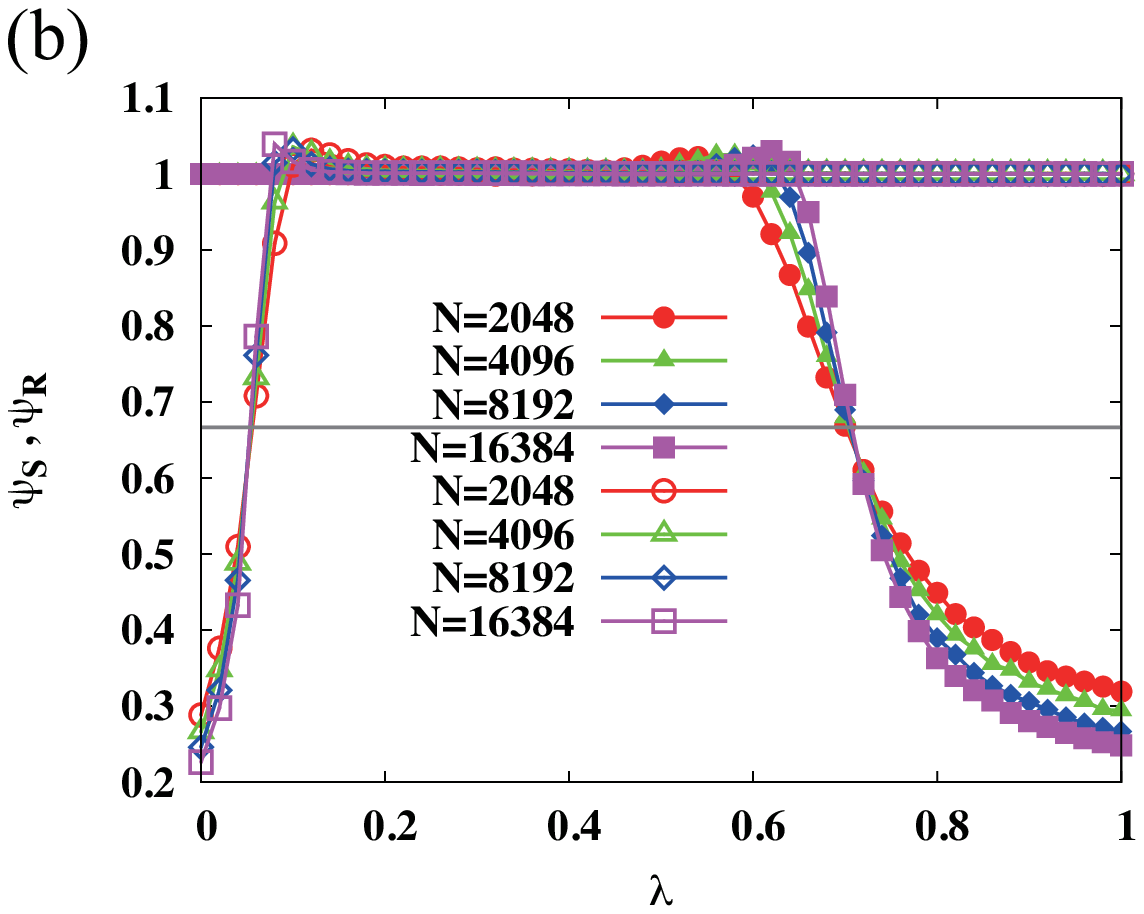}
  }
\resizebox{0.7\columnwidth}{!}{%
  \includegraphics{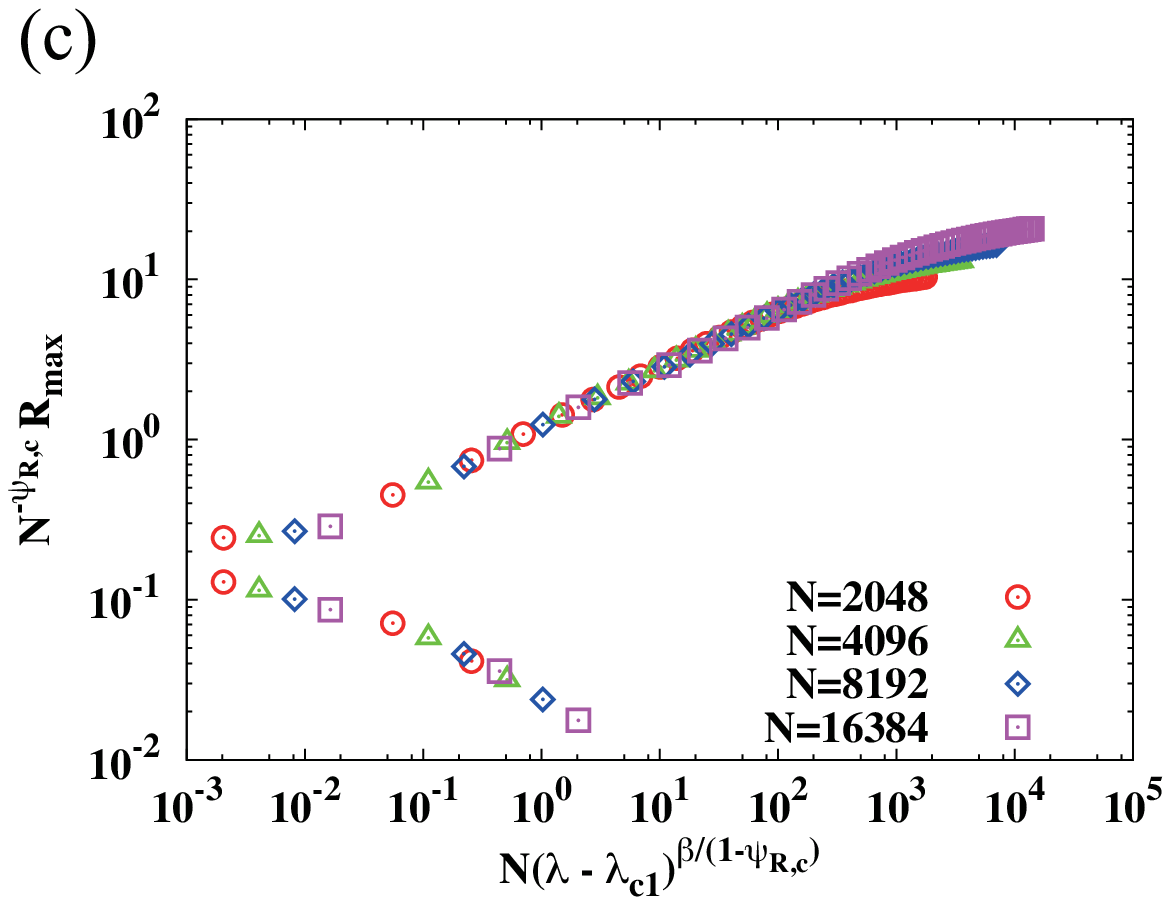}
  }
\resizebox{0.7\columnwidth}{!}{%
  \includegraphics{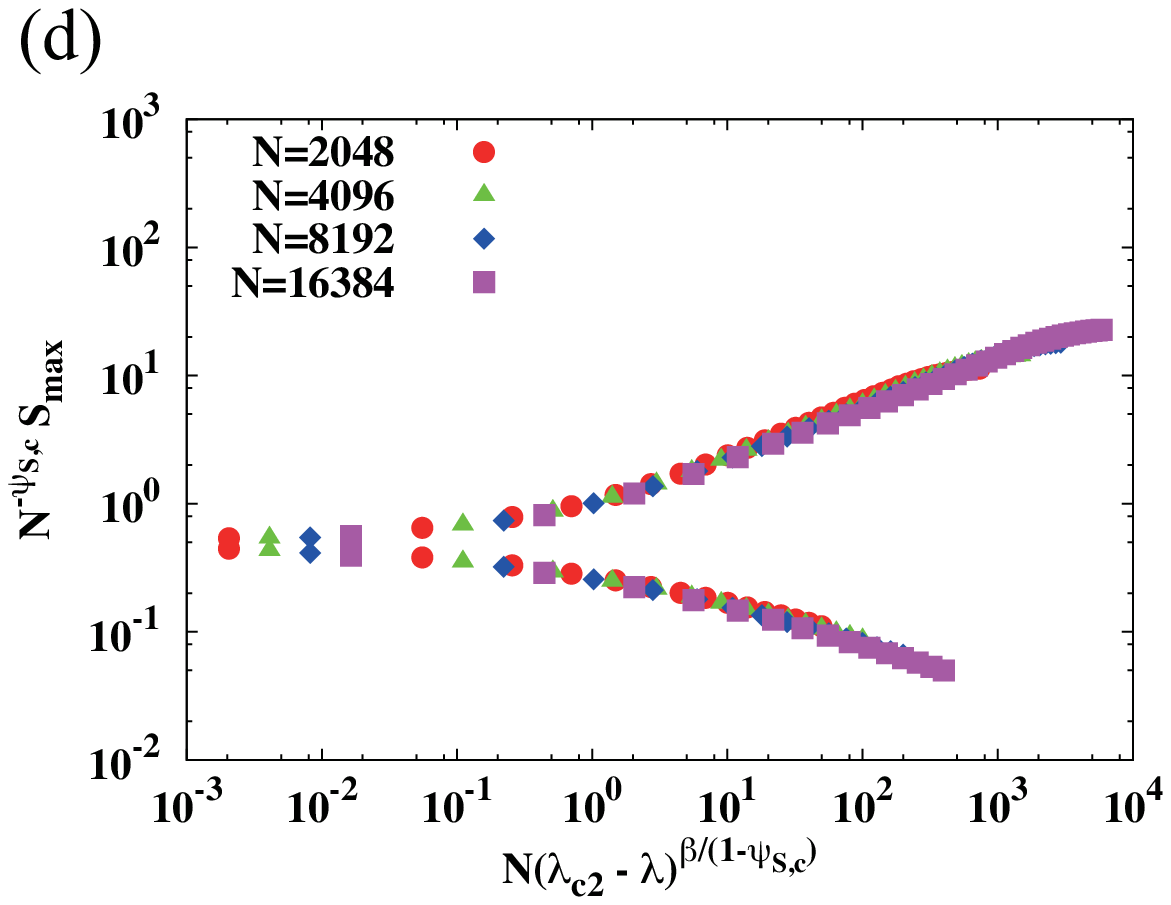}
  }
   \caption{Results for uncorrelated networks with an exponential degree distribution:
(a) order parameters $R_{\rm max}(N)/N$ and $S_{\rm max}(N)/N$, 
(b) fractal exponents $\psi_S(N)$ and $\psi_R(N)$, 
(c) scaling plot of $R_{\rm max}(N)$ and (d) scaling plot of $S_{\rm max}(N)$.
The number of nodes $N$ is 2048 (red-circle), 4096 (green-triangle), 8192 (blue-diamond), and 16384 (purple-square).
The full and open symbols denote the results for the residual and infected networks, respectively.
}
   \label{fig:OrderExp}
\end{figure}


\section{Result: Uncorrelated Case}

In this section, we show the results of Monte-Carlo simulations for the SIR model on uncorrelated networks. 
The number of graph realizations is 500 and the number of SIR runs on each realization is 100.
The number of nodes is taken from $N=2^9$ to $2^{14}$.

First, we consider uncorrelated networks with the exponential degree distribution (\ref{degExp}).
The order parameters $R_{\rm max}(N)/N$ and $S_{\rm max}(N)/N$ with several network sizes 
are shown in Fig.~\ref{fig:OrderExp}(a).
Here we set $p_{\rm{seed}}=0.1$.
As the infection rate $\lambda$ increases, 
$R_{\rm max}(N)/N$ monotonically increases while $S_{\rm max}(N)/N$ monotonically decreases.

To determine the critical points, we introduce the fractal exponents 
$\psi_R$ and $\psi_S$ of the infected and residual networks defined by \cite{nogawa2009monte,hasegawa2010profile}
\begin{eqnarray}
 S_{\rm{max}}(N) \propto N^{\psi_S}, \quad  R_{\rm{max}}(N) \propto N^{\psi_R},
\end{eqnarray}
respectively. 
Noting that Eq.(\ref{Phase1}), we expect 
\begin{eqnarray}
\{\psi_R, \psi_S \} \propto
{\Biggl\{}
\begin{array}{ccl}
\{ 0, 1 \} & (\lambda < \lambda_{c1}) & \\
\{ \psi_{R,c}, 1 \} & (\lambda= \lambda_{c1}) & \\
\{ 1, 1 \} & (\lambda_{c1} <\lambda< \lambda_{c2}) & \\
\{ 1, \psi_{S,c} \} & (\lambda= \lambda_{c2}) & \\
\{ 1,0 \} & (\lambda> \lambda_{c2}) &
\end{array}.
\end{eqnarray}
For numerical computations, we evaluate the effective exponents of finite $N$:
\begin{eqnarray}
 \psi_S(N) = \frac{{\rm{d}}\log S_{\rm{max}}(N)}{{\rm{d}} \log N}, \quad 
 \psi_R(N) = \frac{{\rm{d}}\log R_{\rm{max}}(N)}{{\rm{d}} \log N} . 
\end{eqnarray}
Then, $\psi_R(N)$ and $\psi_S(N)$ with several sizes cross at $\lambda =\lambda_{c1}$ and $\lambda =\lambda_{c2}$, respectively.

By using data of Fig.~\ref{fig:OrderExp}(a), we plot the effective fractal exponent in Fig.~\ref{fig:OrderExp}(b).
We have a crossing point of $\psi_R(N)$ at $\lambda_{c1}\approx 0.05$, and of $\psi_S(N)$ at $\lambda_{c2}\approx 0.72$.
At the crossing points, the corresponding fractal exponents take the same value 
$\psi_{R,c}=\psi_{S,c}=2/3$. 
The property that the largest component size at criticality is proportional to $N^{2/3}$ is also observed in 
the mean field SIR model \cite{kessler2007solution,ben2004size} and percolation of the random graph \cite{janson1993birth}.


\begin{figure}
  \centering
\resizebox{0.7\columnwidth}{!}{%
  \includegraphics{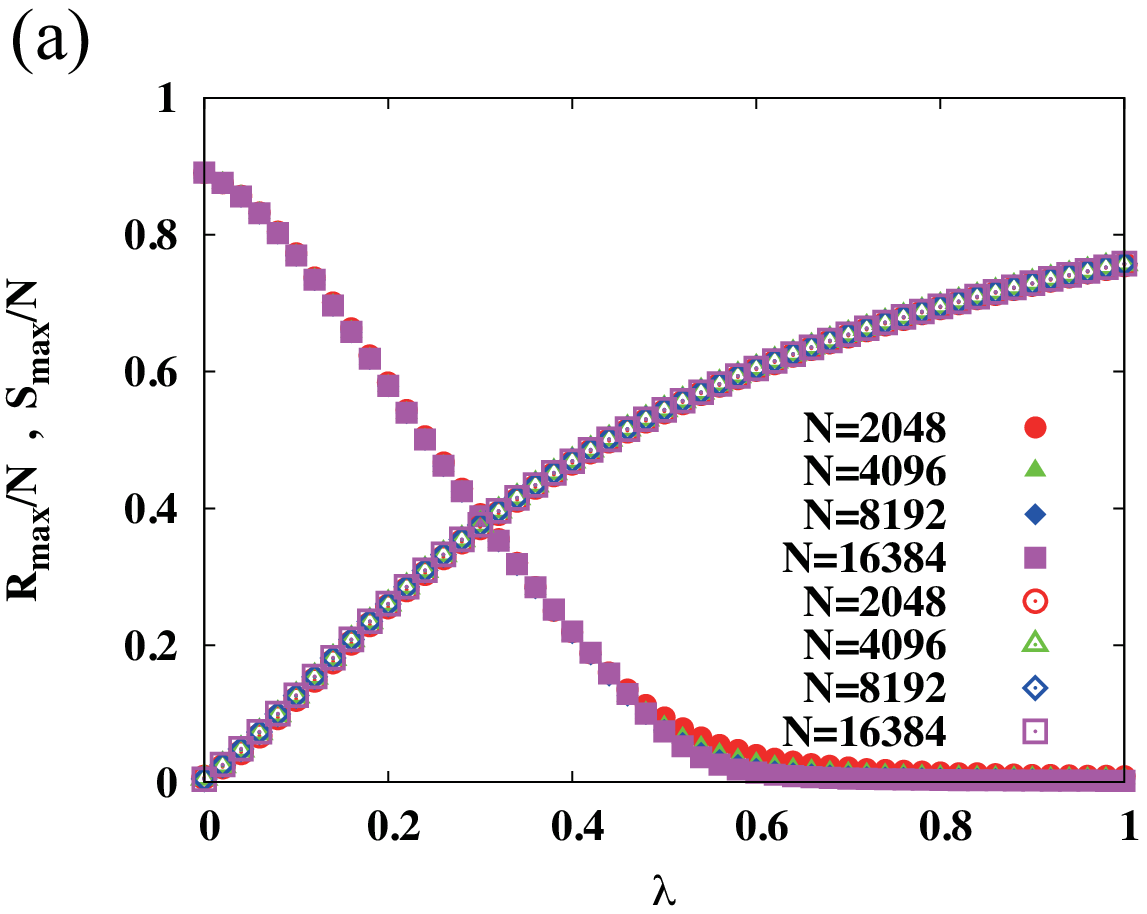}
  }
\resizebox{0.7\columnwidth}{!}{%
  \includegraphics{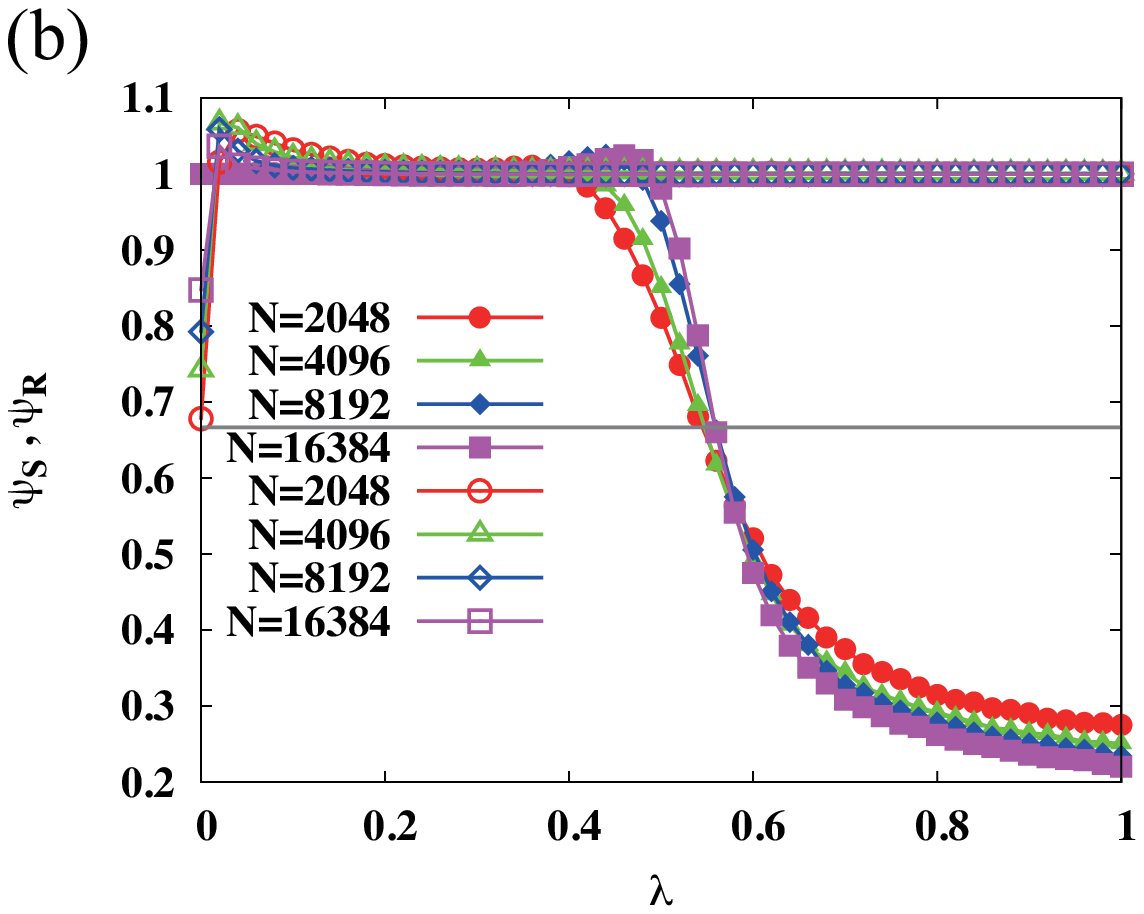}
  }
\resizebox{0.7\columnwidth}{!}{%
  \includegraphics{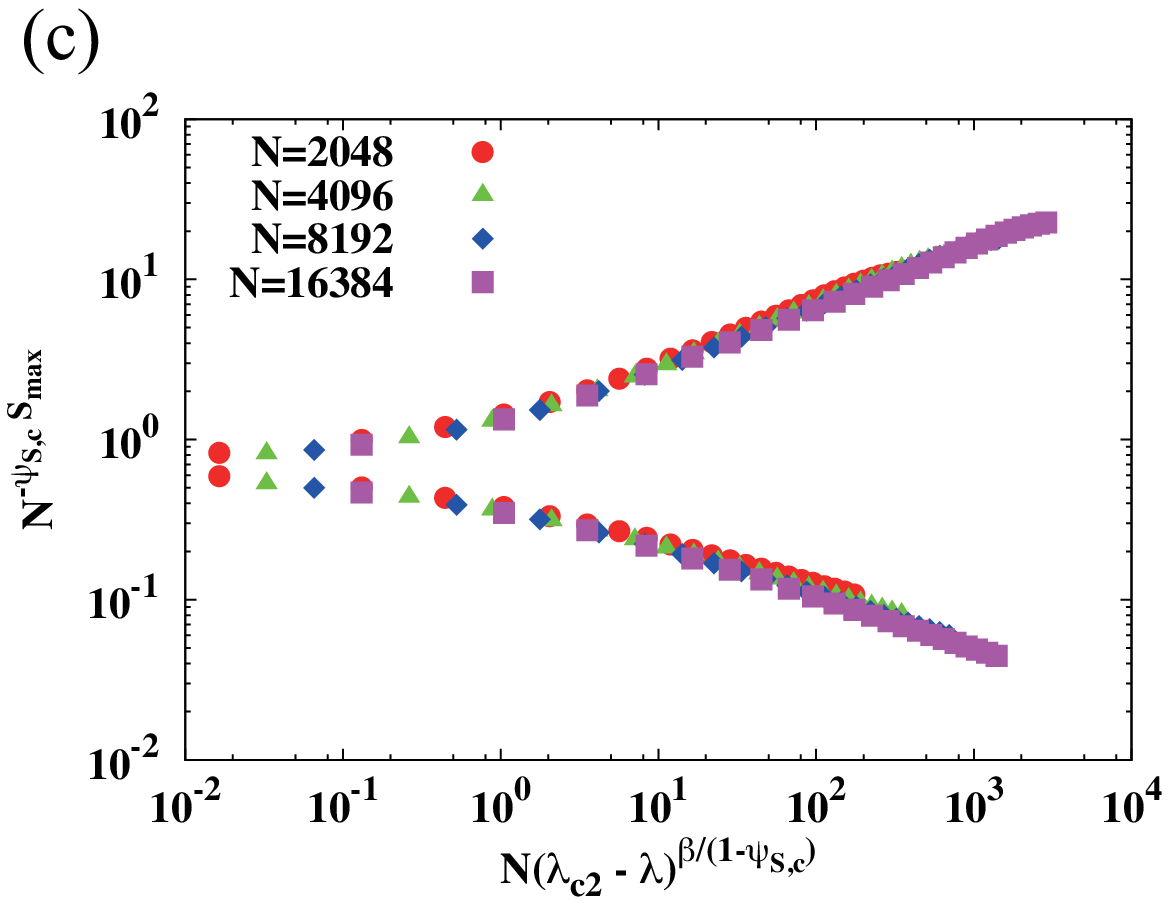}
  }
   \caption{
Results for uncorrelated SF networks:
(a) order parameters $R_{\rm max}(N)/N$ and $S_{\rm max}(N)/N$, 
(b) fractal exponents $\psi_S(N)$ and $\psi_R(N)$, and 
(c) scaling plot of $S_{\rm max}(N)$.
The number of nodes $N$ is 2048 (red-circle), 4096 (green-triangle), 8192 (blue-diamond), and 16384 (purple-square).
The full and open symbols denote the results for the residual and infected networks, respectively.
}
   \label{fig:OrderPow}
\end{figure}


We further determine the critical exponent by using a finite size scaling.
For $R_{\rm max}(N)$, we assume the following scaling form \cite{hasegawa2010profile}: 
\begin{equation}
R_{\rm max}(N) = N^{\psi_{R,c}} f [ N (\Delta \lambda)^{\beta/(1-\psi_{R,c})}] \,, 
\label{SmaxScalingForm-recon}
\end{equation}
where $f(x)$ is a scaling function satisfying
\begin{equation}
f(x) =
{\Biggl\{}
\begin{array}{ccl}
x^{1-\psi_{R,c}} & {\rm for} & x \gg 1 \\
{\rm const} & {\rm for} & x \ll 1 
\end{array}, 
\label{SmaxScaling1}
\end{equation}
$\Delta \lambda=\lambda -\lambda_{c1}$, and $\beta$ is the critical exponent of the order parameter, 
$R_{\rm max}(N)/N \propto |\Delta \lambda|^\beta$.
The scaling form for $S_{\rm max}(N)$ is assumed in a similar way. 
Our scaling results for $R_{\rm max}(N)$ and $S_{\rm max}(N)$ are shown in Figs.~\ref{fig:OrderExp}(c) and (d).
For both plots, 
the fitting parameter $\beta$ is taken as $\beta=1$. 
This indicates that these transitions are in the mean field universality class.
Similar scalings for the mean cluster sizes of the residual and 
infected networks are also assumed in a similar way \cite{hasegawa2010profile}. 
We have a good data collapse for both by setting $\beta =1$ (not shown).

The results for the SF network are shown in Fig.\ref{fig:OrderPow}.
As $N$ increases, $\psi_{R}(N)$ near $\lambda=0$ approaches one, 
indicating that $\lambda_{c1}=0$ for SF networks with $\gamma=3$ \cite{moreno2002epidemic}.
On the other hand, $\psi_{S}(N)$ with several sizes cross at $\lambda_{c2} \approx 0.56$, 
which is smaller than that of the exponential distribution. 
This is consistent with the fact that the SF network with $\gamma=3$ is fragile compared to the random graph with the same average degree~\cite{hasegawa2011robustness}.
The same analysis is performed with several initial seed fractions $p_{\rm seed}=0.3, 0.5, 0.7$. 
Our simulations indicate that the SF network remains fragile in this range of $p_{\rm seed}$.
In Fig.\ref{fig:OrderPow}(c), our scaling result shows that the transition of residual networks belongs to the mean field universality class, 
$\beta=1$, even when the network is scale-free with $\gamma=3$.
Here we assumed $\psi_{S,c}=2/3$. Although our numerical data (Fig.\ref{fig:OrderPow}(b)) seems to be less precise due to small samples, 
we performed a similar analysis for a dynamic percolation in the same universality as the SIR model to show precisely $\psi_{S,c}=2/3$ for that model (not shown).


\begin{figure}[tb]
 \centering
\resizebox{0.7\columnwidth}{!}{%
  \includegraphics{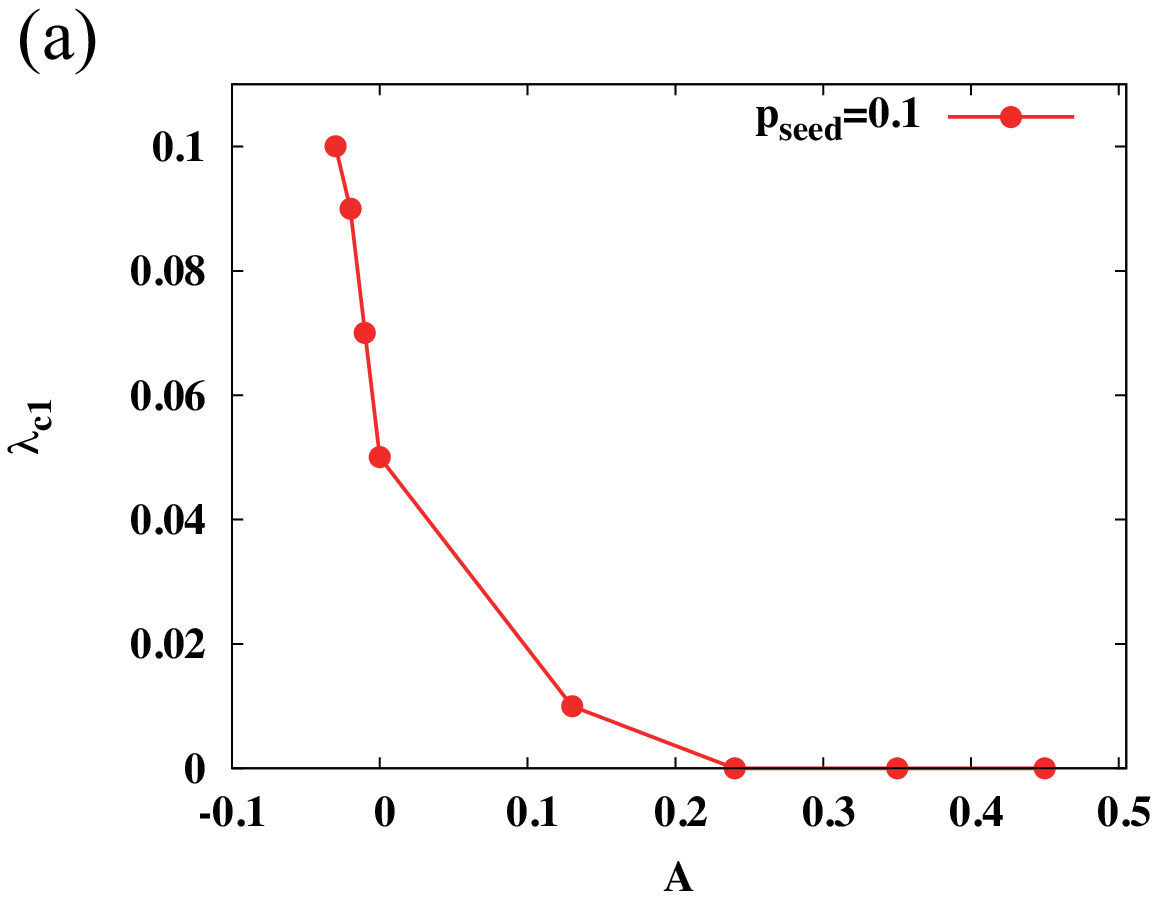}
  }
\resizebox{0.7\columnwidth}{!}{%
  \includegraphics{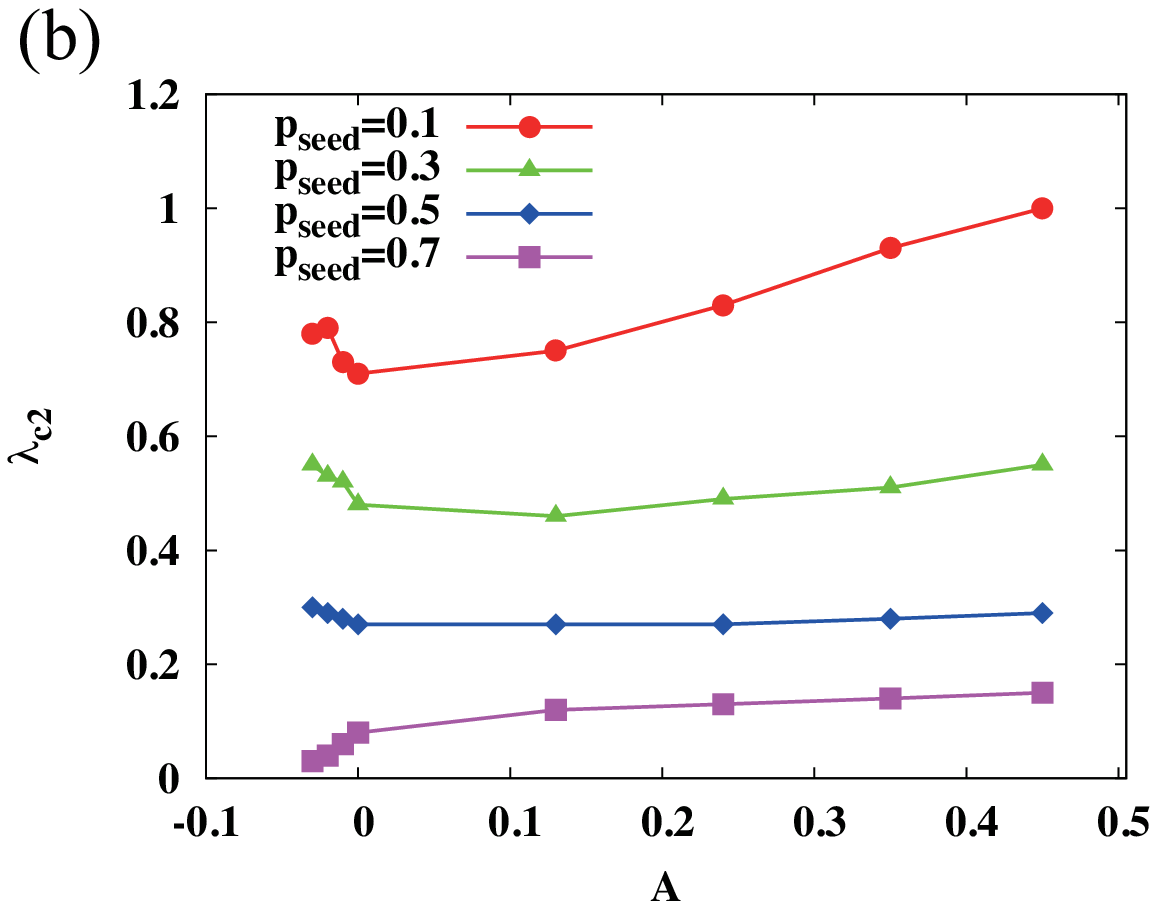}
  }
 \caption{
(a) $\lambda_{c1}$ and (b) $\lambda_{c2}$ 
of the correlated network with an exponential degree distribution as a function of degree correlation $\mathcal{A}$.
The initial seed fraction $p_{\rm seed}$ is set to 0.1 (red-circle), 0.3 (green-triangle), 0.5 (blue-diamond), and 0.7 (purple-square).
}
 \label{fig:CorLambdac1Exp}
\end{figure}


\begin{figure}[t]
 \centering
\resizebox{0.7\columnwidth}{!}{%
  \includegraphics{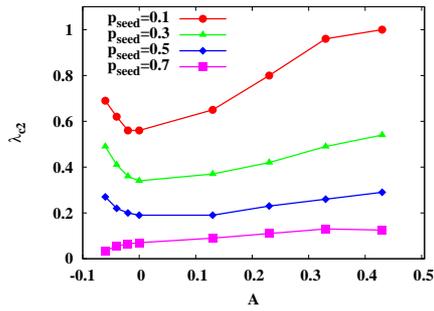}
  }
 \caption{
$\lambda_{c2}$ of the correlated network with the SF degree distribution 
as a function of degree correlation $\mathcal{A}$.
The initial seed fraction $p_{\rm seed}$ is set to 0.1 (red-circle), 0.3 (green-triangle), 0.5 (blue-diamond), and 0.7 (purple-square).
}
 \label{fig:CorLambdac1Pow}
\end{figure}

\section{Result: Correlated Case}

We performed similar simulations on correlated networks with several degree correlation $\mathcal{A}$.
Here $\mathcal{A}$ takes a value in a range $[-0.1, 0.5]$. 

The critical infection rates $\lambda_{c1}$ and $\lambda_{c2}$ 
of the network with the exponential degree distribution 
are plotted as a function of $\mathcal{A}$ (Fig.\ref{fig:CorLambdac1Exp}).
As expected in \cite{vazquez2003resilience}, 
$\lambda_{c1}$ decreases when $\mathcal{A}$ increases 
(Fig.\ref{fig:CorLambdac1Exp}(a)).
On the other hand, we observe a nontrivial behavior for $\lambda_{c2}$ (Fig.\ref{fig:CorLambdac1Exp}(b)). 
For $p_{\rm{seed}}=0.1, 0.3, 0.5$, the uncorrelated network is most fragile: 
$\lambda_{c2}$ increases when the network is correlated, 
regardless of whether it is assortative or disassortative.
This tendency decreases with the increase of $p_{\rm{seed}}$, 
and we did not observe such a behavior for $p_{\rm{seed}}=0.7$. 
There, a disassortative network is fragile than the uncorrelated one.
Figure \ref{fig:CorLambdac1Pow} is the result for SF network. 
We can observe a similar dependence of $\lambda_{c2}$ on $\mathcal{A}$.
Our scaling results indicate that 
$\psi_{S,c}$ (and $\beta$) tends to decrease (increase) with the increase of the assortativity, 
although it was hard for our simulation to determine precise values because of large error-bars (not shown). 
More extensive simulation or another effective method should be performed for the evaluation of the critical properties in detail.

\begin{figure*}[!tbp]
 \begin{center}
\resizebox{0.6\columnwidth}{!}{%
  \includegraphics{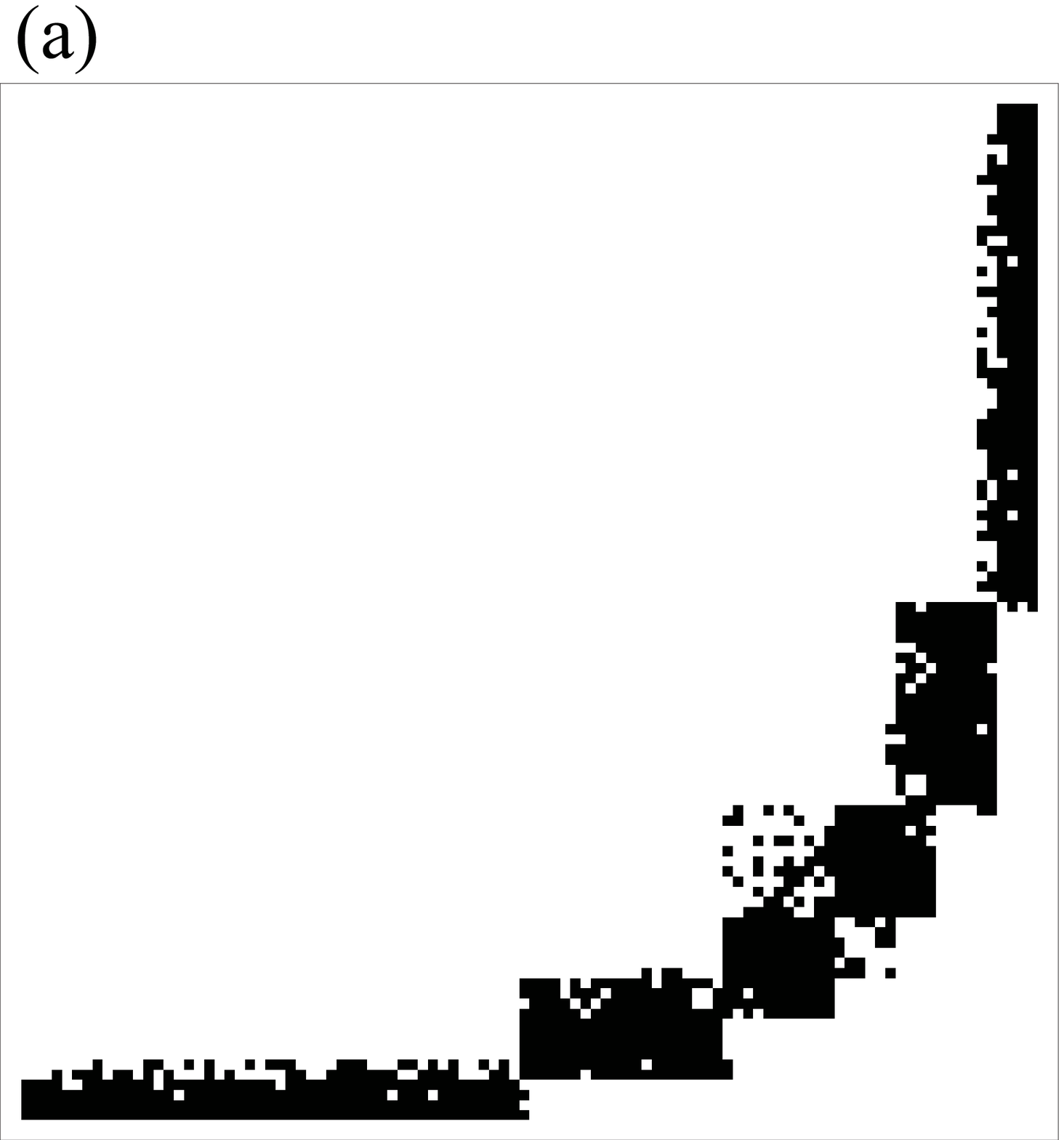}
  }
\resizebox{0.6\columnwidth}{!}{%
  \includegraphics{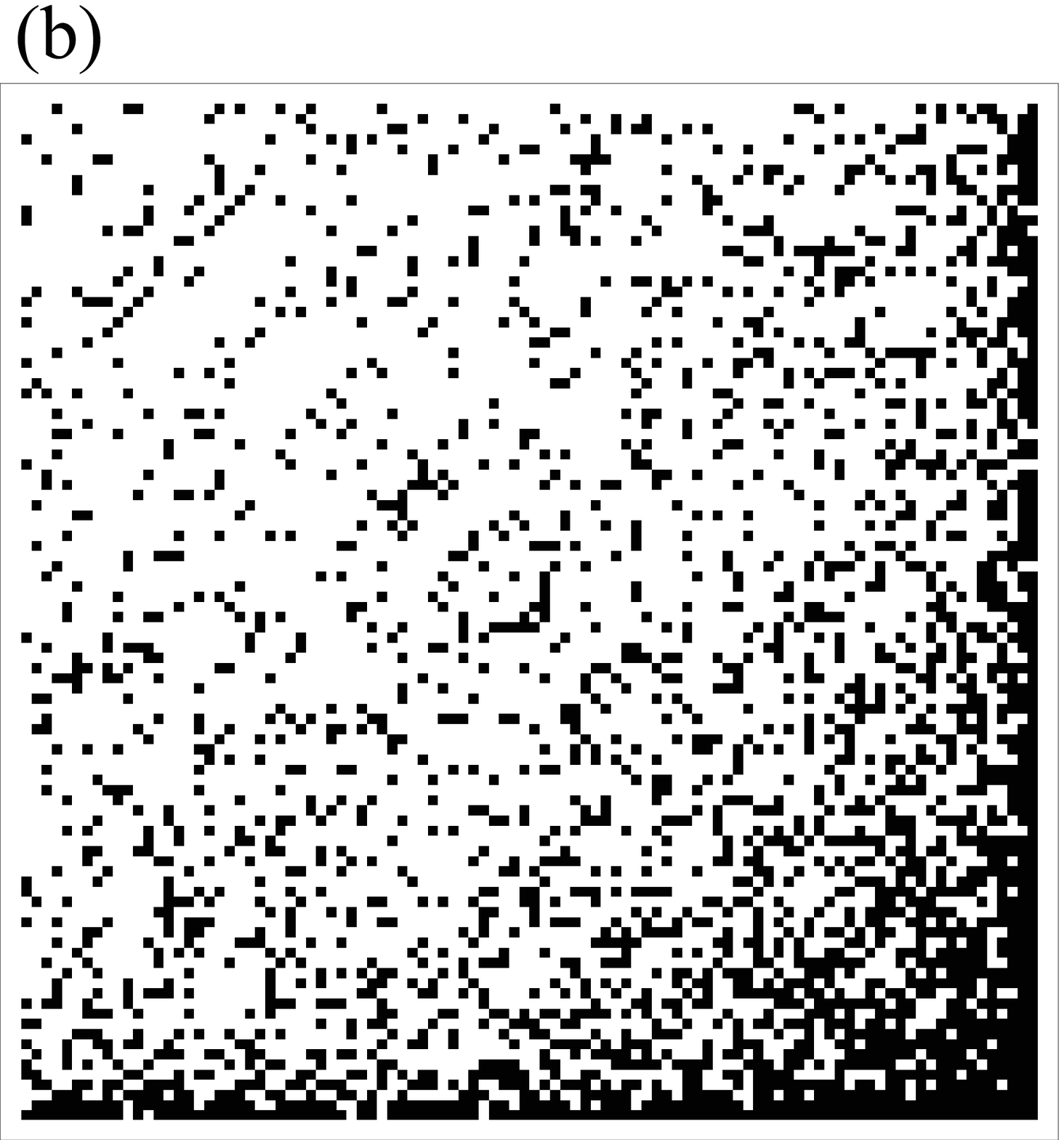}
}
\resizebox{0.6\columnwidth}{!}{%
  \includegraphics{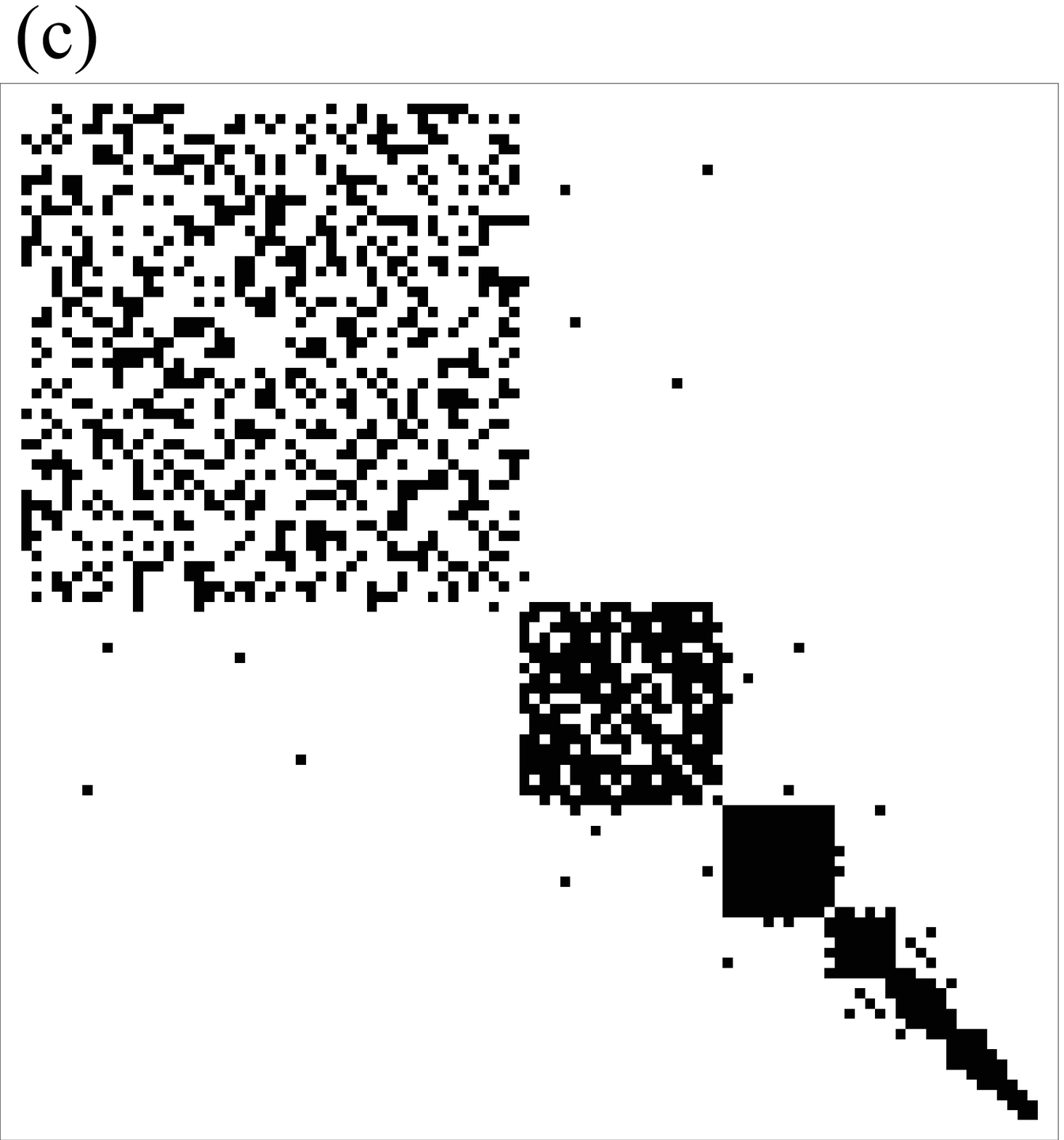}
}
 \end{center}
 \caption{
The ordered adjacency matrix ${\bf A}$ of a SF network with 
(a) $\mathcal{A} \simeq -0.1$, (b) $\mathcal{A} \simeq 0$, and (c) $\mathcal{A} \simeq 0.8$.
The number of nodes is $N=1000$.
The matrices have $N \times N$ entries $A_{ij}$, 
where $A_{ij} = 1$ (black) if nodes $i$ and $j$ are connected and $A_{ij} = 0$ (white) otherwise. 
The nodes are arranged in the order of increasing degree.
 }
 \label{fig:AdjacencyMatrix}
\end{figure*}

The above behavior of $\lambda_{c2}$ is due to the layered network structure of correlated networks \cite{menche2011sequences}.
In Fig.~\ref{fig:AdjacencyMatrix}, we profile the adjacency matrix of SF network for several $\mathcal{A}$.
Assortative networks consist of monolayers of nodes where nodes 
with similar degrees are connected to each other, 
while disassortative networks consist of bilayers where nodes with the low-degree side are connected to a range of high-degree nodes. 
For such a layered structure, there are few edges between different layers.


Our propagating attack consists of two elements: the random initial failure realized by $p_{\rm seed}$ and the SIR dynamics starting from infected nodes.
When the SIR dynamics starts from a single infected node, 
the outbreak from the node, local or global, is likely to be confined in the layer containing the node and
all other layers will remain as the giant component after the attack.
Therefore, correlated networks, assortative or disassortative, are robust compared to uncorrelated ones for small $p_{\rm seed}$. 
For large $p_{\rm seed}$, on the other hand, 
the initial failure damages a number of susceptive nodes and
the remaining network becomes marginally percolating at the beginning.
In this situation the more assortative (disassortative) is the correlation, the more robust (fragile) is 
the network against the random failure, just as seen in the percolation on correlated networks \cite{vazquez2003resilience}.

\section{Summary}

We have investigated the robustness of correlated networks against a propagating attack modeled by the SIR model. 
Our numerical results show that 
correlated networks are robust compared to the uncorrelated ones, 
regardless of whether they are assortative or disassortative when a fraction of infected nodes in an initial state is not too large. 
This behavior is related to the layered structure of our correlated networks.
For large $p_{\rm seed}$, disassortative network becomes fragile.
Note that the layered network structure is not an intrinsic property of correlated networks 
(e.g., we can construct a disassortative network having almost no layered structure \cite{goltsev2008percolation}). 
It remains open for future works to search a mechanism of 
strengthening the robustness of real correlated networks against propagating attacks.

Previous studies for epidemics on complex networks have mainly focused on the (first) critical infection rate $\lambda_{c1}$, above which a global outbreak occurs.
Instead, we have focused on the second critical infection rate $\lambda_{c2}$ in this paper.
In particular, it is important when we consider epidemics on community networks, 
where we should distinguish local epidemics, i.e., an outbreak confined in a single community, and global epidemics, 
i.e., epidemic spreading through the whole communities \cite{masuda2009immunization}.
We expect that the robustness against propagating attacks will be a good measure for searching effective vaccines in community networks.

\section*{Acknowledgements}
TH thanks Toshihiro Tanizawa for helpful comments. 
TH acknowledges the support provided by the Japan Society for the Promotion of Science through Grant-in-Aid for Young Scientists (B) (Grant No. 24740054).

%
%
%
%

\end{document}